# Layer-Dependent Phonons, Excitons, and Magneto-Optical Phenomena in CrSBr: A Mini Review

*Muhammad Aftab[1,2†], Chinmay Kumar Mohanty[3,4†], Zain Ashfaq[5], Warisha Mehmood[1,2], Xiaoli Wang[1,2], Clément Faugeras[4], Wajid Ali[3*], Maciej R. Molas[3]**

[1] CAS Key Laboratory of Nanosystem and Hierarchical Fabrication, National Center for Nanoscience and Technology, Beijing 100190, P.R. China

[2] School of Nanoscience and Engineering, University of Chinese Academy of Sciences, Beijing 100049, P. R. China

[3] Institute of Experimental Physics, Faculty of Physics, University of Warsaw, Pasteura 5, 02-093 Warsaw, Poland

[4] Laboratoire National des Champs Magnétiques Intenses, CNRS–UGA–UPS–INSA–EMFL, 38042 Grenoble, France

[5] Department of Physics, University of Sapienza, Rome, Italy

Corresponding authors*: wajid.ali@fuw.edu.pl, maciej.molas@fuw.edu.pl

**Abstract**

Two-dimensional layered magnetic materials offer a versatile platform for exploring low-dimensional magnetism and coupled many-body interactions in these materials. Chromium sulfur bromide (CrSBr) is a promising candidate for advanced spintronic and optoelectronic applications because of its intrinsic air stability, semiconducting nature, strong in-plane anisotropy, and A-type antiferromagnetic ordering. This review summarizes recent advances in the understanding of the layer-dependent vibrational and excitonic properties of CrSBr, as well as its magneto-optical response, from bulk crystals to the monolayer limit. We examined its crystal structure, magnetic anisotropy, and interlayer spin reorientation, followed by insights into vibrational dynamics and spin-phonon coupling. Particular emphasis is placed on the excitonic landscape, including magnetic-field-sensitive photoluminescence, localized excitonic states, and the coexistence of Frenkel- and Wannier-Mott excitons in the bandgap. Finally, we discuss the challenges and prospects of harnessing the unique layer-dependent properties of CrSBr in spintronic, magneto-optical, and quantum photonic technologies.

## 1. Introduction

Engineering magnetism and excitonic phenomena through dimensionality control in van der Waals (vdW) solids is a rapidly advancing field in condensed matter physics. Reducing the crystal thickness from bulk to few-layer and monolayer forms systematically modulates the interactions among the spin, charge, lattice, and excitonic degrees of freedom, enabling access to emergent quantum states and many-body effects beyond thermodynamic equilibrium [1-4]. Two-dimensional magnetic materials have catalyzed transformative advances in multiple technological domains, including spin-filter tunneling, magnetic memory architectures, valleytronics platforms, excitonic engineering, and quantum photonics[3, 5]. This is fundamentally because their reduced dimensionality amplifies the role of interfacial and interplane interactions while simultaneously suppressing competing perturbations that typically mask quantum effects in bulk materials [1, 3, 6-11].

Among these, chromium sulfur bromide has emerged as a distinguished system owing to its intrinsic air stability, semiconducting nature, pronounced in-plane anisotropy, and robust A-type antiferromagnetic ordering with a Néel temperature near 132 K. Unlike chromium trihalides ($CrX_3$, where X = Cl, Br, or I), which often suffer from low ordering temperatures and environmental instability [12-15]. CrSBr combines practical stability, strong magnetic anisotropy, and a quasi-one-dimensional crystal structure. This unique combination results in highly anisotropic magnetic and optical properties that evolve systematically with the number of layers [16-19]. Recent experimental and theoretical advances have elucidated the layer-dependent evolution of the magnetism, vibrational dynamics, and excitonic landscape of CrSBr [7, 20-22].

Key phenomena include interlayer spin reorientation, strong spin-phonon coupling, and the coexistence of Frenkel and Wannier–Mott excitons, which collectively link dimensionality to coupled magnetic and optical responses [16-19, 23-25]. Therefore, this mini review critically examines recent progress in understanding the crystal structure, magnetic anisotropy, interlayer magnetic interactions, vibrational properties, and excitonic behavior of CrSBr, from bulk crystals to monolayer limits. We emphasize the fundamental mechanisms driving layer-dependent magneto-optical responses and their implications for spintronic, magneto-optical, and quantum photonic applications.

## 2. Crystal Structure and Structural Anisotropy

Chromium sulfur bromide crystallizes in an orthorhombic FeOCl-type structure (space group Pmmn), featuring weakly coupled layers stacked along the crystallographic c-axis via van der Waals (vdW) interactions **(Figure 1a)** [26]. Each monolayer consists of distorted $CrS_4Br_2$ octahedra arranged in a highly anisotropic rectangular lattice. These octahedra are connected through edge-sharing along the a-axis and corner-sharing along the b-axis, establishing an intrinsically anisotropic bonding environment that strongly influences the electronic, magnetic, vibrational, and optical properties of the material [27]. The $Cr^{3+}$ ions coordinate with four $S^{2-}$ and two $Br^{-}$ anions, with Br atoms occupying the outermost atomic layers. These bromine atoms play a critical role in mediating the interlayer coupling and vibrational interactions. The vdW gaps between adjacent layers facilitate mechanical exfoliation down to the monolayer limit while preserving the lattice symmetry [26, 28]. First-principles density functional theory (DFT) calculations, supported by low-temperature X-ray and neutron diffraction experiments, confirmed the structural stability and lattice parameters (a = 3.52 Å, b = 4.74 Å, c = 7.91 Å) of CrSBr, which closely matched the experimental values [29-31]. The weak vdW interlayer forces ensure that the few-layer and monolayer samples retain the bulk crystallographic framework, preserving the pronounced anisotropy essential for the observed layer-dependent phenomena [29]. Beyond the experimentally confirmed AA-0 stacking, theoretical studies have identified several high-symmetry bilayer configurations, namely, AA-1, AB-0, and AB-1 **(Figure 1b–e)**, which differ in terms of their relative lateral displacements and symmetry operations [26]. Despite these variations, DFT consistently predicts the A-type antiferromagnetic (AFM) state as the energetically favored magnetic configuration, characterized by ferromagnetic spin alignment within layers coupled antiferromagnetically between adjacent layers, resulting in alternating spin orientations along the c-axis [32]. This robustness highlights the intrinsic magnetic stability of CrSBr and its suitability for low-dimensional spintronic applications [32, 33].

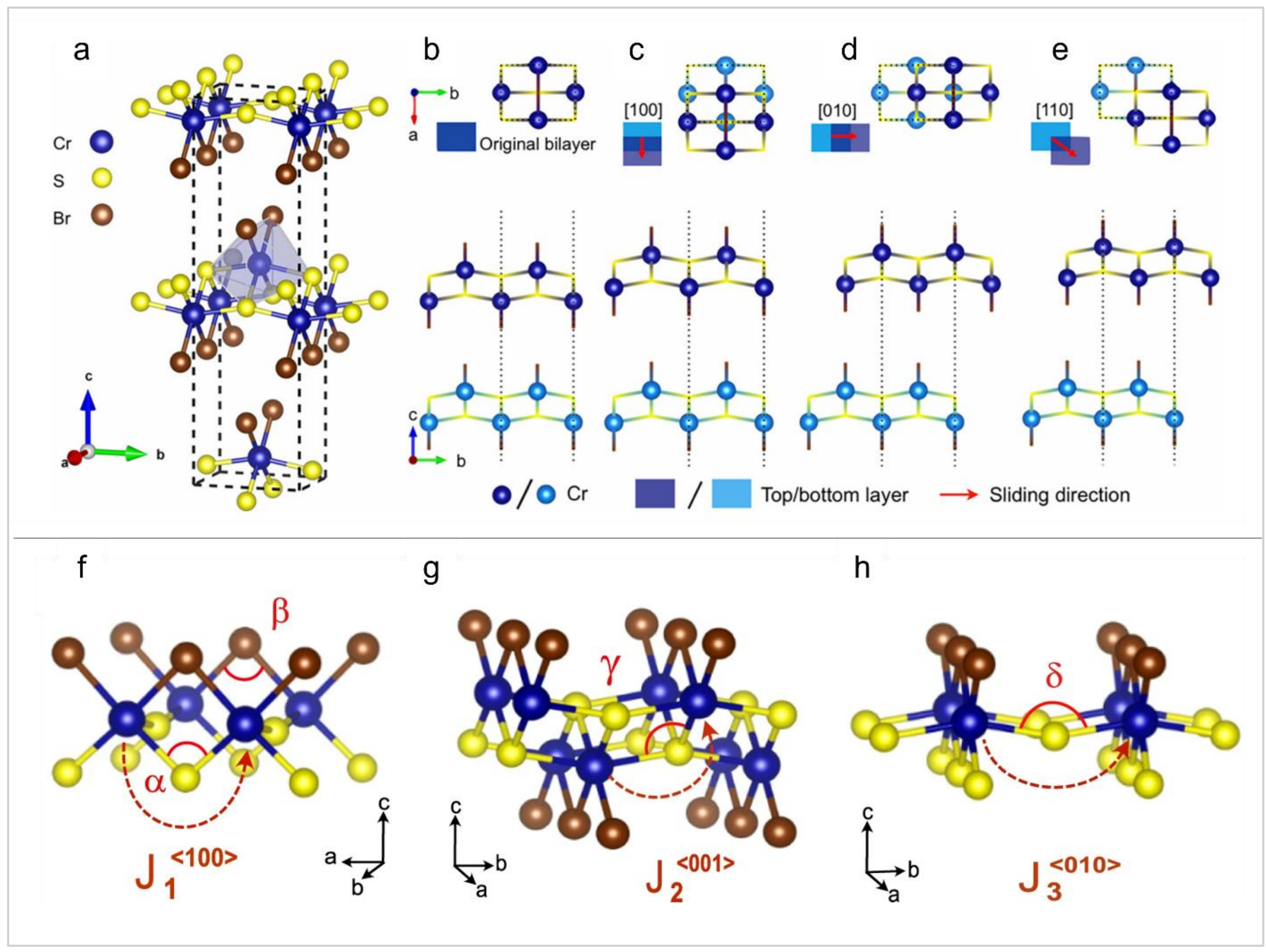

**Figure 1.** (a) illustrates the crystal structure of bulk CrSBr. (b)–(e) Top and side perspectives of the four CrSBr two-layer (2 L) structures, each exhibiting distinct stacking configurations: AA-0, AA-1, AB-0, and AB-1. For simplicity, the S and Br atoms were omitted from the 2 L structures. The rectangles, colored light and dark blue, denote the bottom and top layers, respectively. The red arrows indicate the sliding directions of the CrSBr layer along the [100], [010], and [110] axes of the 2Ls. Reproduced with permission. [26] Copyright 2025, American Physical Society. (f-h) emphasize the three primary intralayer superexchange pathways ((J1), (J2), and (J3)) along with their respective bond angles (α, β, γ, and δ). Reproduced with permission. [29] Copyright 2022, Nature Springer.

The microscopic origin of the magnetic anisotropy and interlayer coupling arises from multiple intralayer super-exchange pathways mediated by the sulfur and bromine atoms **(Figure 1f–h)** [29]. These pathways, denoted $J_1$, $J_2$, and $J_3$, are defined by distinct Cr–S/Br–Cr bond angles (α, β, γ, and δ) and specific orbital hybridizations, governing the balance between ferromagnetic intralayer and antiferromagnetic interlayer interactions. The edge-sharing geometry along the a-axis facilitates a strong ferromagnetic superexchange via the direct overlap of Cr 3d orbitals mediated predominantly by S 3p orbitals, forming quasi-one-dimensional magnetic chains. In contrast, corner-sharing connectivity along the b-axis introduces an anisotropic exchange coupling, which significantly contributes to the pronounced in-plane magnetic anisotropy. Bromine atoms positioned in the outermost layers mediate superexchange pathways that modulate the interlayer magnetic coupling and vibrational interactions. This bromine-mediated superexchange plays a pivotal role in linking structural anisotropy to magnetic ordering and magneto-optical response. The interplay among these exchange pathways underpins the stability of the A-type AFM ground state and governs the layer-dependent evolution of the spin reorientation, spin-phonon coupling, and excitonic phenomena characteristic of CrSBr [29, 31, 34].

## 3. Layer-Dependent Interlayer Magnetism and Spin Reorientation

A-type AFM ordering in CrSBr persists down to the 2L limit, establishing it as a prototypical system for investigating thickness-dependent magnetism in two-dimensional materials [35-38]. First-principles calculations have revealed that monolayer CrSBr favors a ferromagnetic ground state, whereas multilayer structures stabilize an A-type AFM configuration with the magnetic easy axis aligned along the crystallographic b-axis [18, 39]. Thus, increasing the number of layers beyond the monolayer limit drives a transition to the A-type AFM state, which is characterized by ferromagnetic intralayer spin alignment and antiferromagnetic coupling between adjacent layers. This evolution highlights the crucial role of interlayer exchange interactions in stabilizing the magnetic ground state of the multilayers. As depicted in **Figure 2a** and **2b**, both the nearest-neighbor interlayer exchange interaction (J) and magnetic anisotropy energy (K) strongly depend on the layer number. Importantly, these parameters exhibited an odd-even oscillatory pattern, reflecting the interplay between magnetic anisotropy, interlayer exchange coupling, and Zeeman energy under applied magnetic fields [18].

Applying a magnetic field along the easy b-axis induces a spin-flop transition governed by the competition among the antiferromagnetic interlayer exchange (which favors antiparallel spins), magnetic anisotropy energy (which stabilizes spins along the easy axis), and Zeeman energy (which tends to align spins with the field) [40, 41]. The system undergoes a transition from a collinear

AFM state to a noncollinear spin configuration once the Zeeman energy exceeds the anisotropy energy and approaches the interlayer exchange interaction energy. Both the critical field (Hc), which marks the onset of the spin-flop, and the saturation field (Hs), which corresponds to the full spin alignment, depend strongly on the number of layers **(Figure 2c)**. This behavior demonstrates the tunability of the interlayer magnetic coupling in atomically thin CrSBr [18].

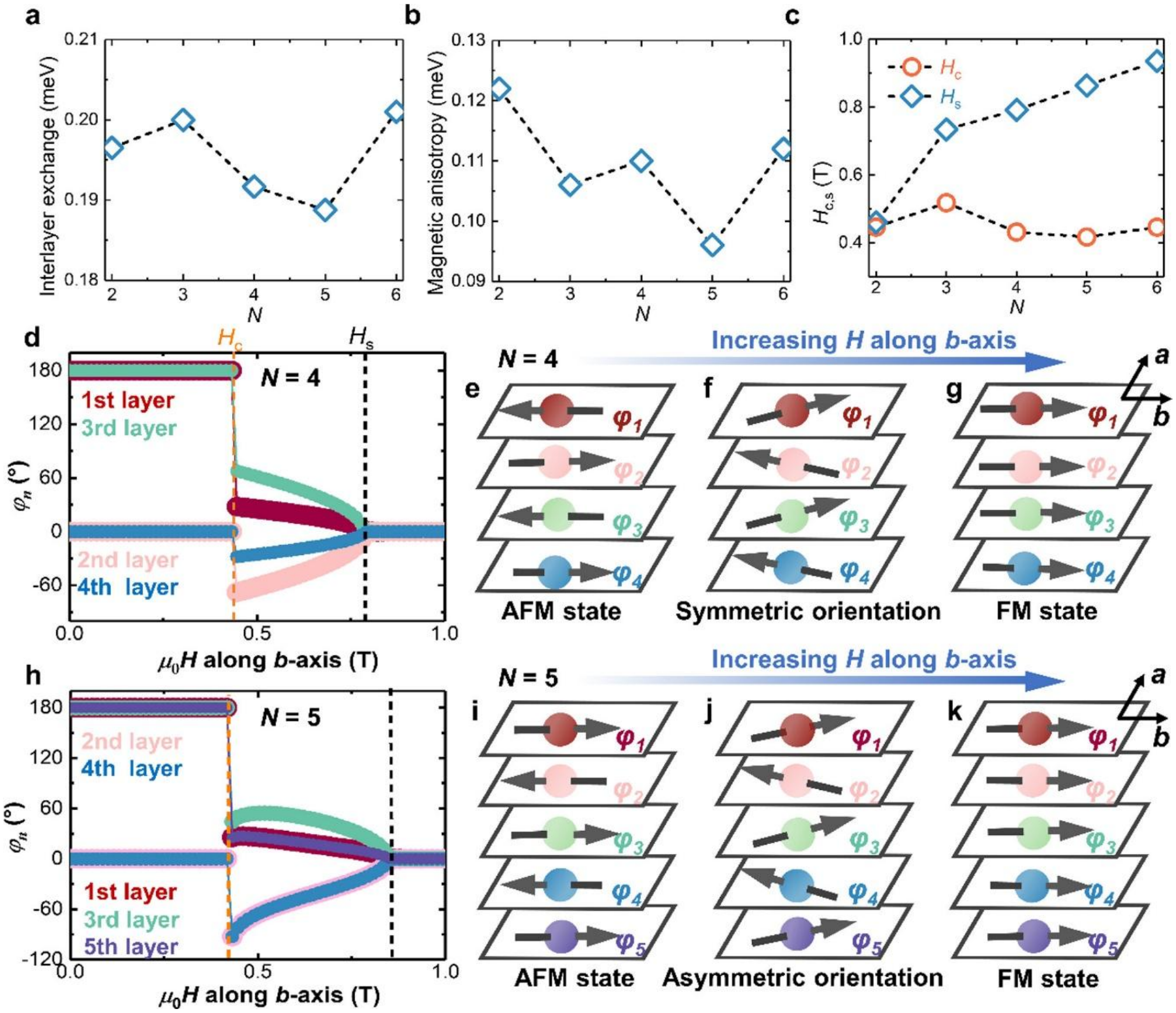

**Figure 2.** Interlayer spin reorientation of CrSBr multilayers: (a) Layer-dependent interlayer exchange energy. (b) Magnetic anisotropy energy vs. layer number, N. (c) Theoretical critical (HcDFT) and saturation (HsDFT) fields of spin-flop transitions. (d, h) Magnetic phase diagrams of tetralayer (N=4, d) and pentalayer (N=5, h) from the AFM linear-chain model showing spin reorientation under b-axis magnetic field. (e−g) Schematic of tetralayer CrSBr spin reorientation from AFM to noncollinear intermediate to FM state. (i−k) Similar schematic highlighting asymmetry in odd-N multilayers due to uncompensated magnetization in noncollinear states. Reproduced with permission.[18] Copyright 2022 American Chemical Society.

Monte Carlo simulations employing an antiferromagnetic linear-chain model successfully reproduce the experimentally observed magnetic phase diagrams and reveal distinct spin-reorientation behaviors between even- and odd-layer CrSBr samples **(Figure 2d–k)** [18]. In the even-layer samples, the intermediate spin-flop phase exhibited a symmetric noncollinear configuration, with opposing layers rotating by equal angles relative to the magnetic field direction. Conversely, odd-layer crystals possess uncompensated magnetization, resulting in an asymmetric noncollinear spin texture. This asymmetry causes the equivalent layers to adopt inequivalent spin orientations, thereby breaking the spin symmetry in a thickness-dependent manner. This odd-even effect correlates with the experimentally observed differences in the magnetotransport and magnetic responses of ultrathin CrSBr.

The spin reorientation that varies with the layer thickness highlights the subtle balance among the interlayer exchange interactions, magnetic anisotropy, and Zeeman energy, which collectively shape the magnetic phase diagram of CrSBr. The appearance of unique magnetic configurations in samples with odd and even numbers of layers illustrates how dimensionality significantly affects both the magnetic ground state and spin dynamics influenced by external fields. These magnetic phases, which depend on the thickness, provide a fundamental basis for comprehending the interaction between the magnetic order and other degrees of freedom, such as lattice vibrations and optical excitations, which are discussed in the following sections.

## 4. Vibrational Properties and Spin–Phonon Coupling in CrSBr

The coupling between lattice vibrations and magnetic order is a fundamental characteristic of low-dimensional magnetic materials and critically influences their electronic and optical properties[42-47]. In CrSBr, this interplay is especially pronounced owing to its quasi-one-dimensional crystal structure, strong in-plane anisotropy, and layered antiferromagnetic ground state. The reduced dimensionality and anisotropic bonding environment result in highly directional vibrational, electronic, and optical responses, positioning CrSBr as an excellent platform for studying the interactions among the spin, lattice, electronic, and excitonic degrees of freedom [48, 49]. Raman scattering spectroscopy is a particularly effective tool for probing these interactions because of its sensitivity to crystal symmetry, magnetic order, and electron–phonon coupling [50-52].

Early evidence of strong spin-phonon coupling in CrSBr was revealed by temperature-dependent Raman measurements performed across an antiferromagnetic transition. Upon cooling below the Néel temperature ($T_N$ ≈132 K), several Raman-active phonon modes displayed anomalous frequency shifts, linewidth variations, and intensity changes that cannot be attributed solely to anharmonic lattice effects **(Figure 3 a and 3b)** [48]. Additionally, new Raman features appeared below $T_N$, indicating crystal symmetry modifications associated with the onset of magnetic ordering. These findings demonstrate a strong coupling between the lattice dynamics and antiferromagnetic correlations. Ab initio phonon calculations successfully replicated the experimental spectra and confirmed that the magnetic phase significantly affected the vibrational response of the spectra [53]. Consequently, the resulting Raman fingerprints serve as sensitive spectroscopic markers for differentiating between magnetic phases and monitoring the evolution of the magnetic order [53-55].

The vibrational spectrum of CrSBr strongly depends on the layer thickness. Raman scattering studies, performed with excitation energies in the vicinity of the electronic transitions of CrSBr, revealed systematic shifts in phonon frequencies and changes in their intensities in samples ranging from monolayers to bulk [54, 55]. Specifically, $A_g^1$ and $A_g^2$ modes exhibited gradual redshifts as the number of layers decreased, whereas the $A_g^3$ mode remained almost constant. This contrasting behavior reflects the anisotropic nature of lattice vibrations and weak interlayer coupling, which are typical of van der Waals materials. Because bromine atoms reside in the outermost atomic layers, they mediate interlayer vibrational interactions that significantly influence the phonon spectrum. As the thickness increased, the restoring forces associated with certain lattice vibrations increased, causing the observed stiffening of the vibrational modes in the multilayer samples [56, 57]. Moreover, Raman measurements performed under excitation energies resonant with CrSBr excitonic transitions revealed additional defect-related vibrational features, underscoring the sensitivity of the phonon spectrum to local structural perturbations and the local magnetic environment (**Figure 3c and 3d**) [55].

Recent polarization-resolved Raman studies have further elucidated the anisotropic nature of the vibrational response of CrSBr. Notably, the $A_g^2$ mode near 244 $cm^{-1}$ exhibited an approximately 90° polarization rotation as a function of the excitation energy and layer thickness [58]. While the $A_g^1$ and $A_g^3$ Modes remain predominantly polarized along the crystallographic b-axis, the $A_g^2$ mode rotates between the a- and b-axes depending on the excitation wavelength and can be represented in **Figure 3e and 3f**. This behavior cannot be solely attributed to lattice symmetry and instead reflects the impact of anisotropic electronic transitions on the Raman scattering process [59, 60]. The pronounced excitation energy dependence of the Raman anisotropy implicates resonant excitonic states, indicating that the vibrational response of CrSBr is closely coupled to its electronic structure.

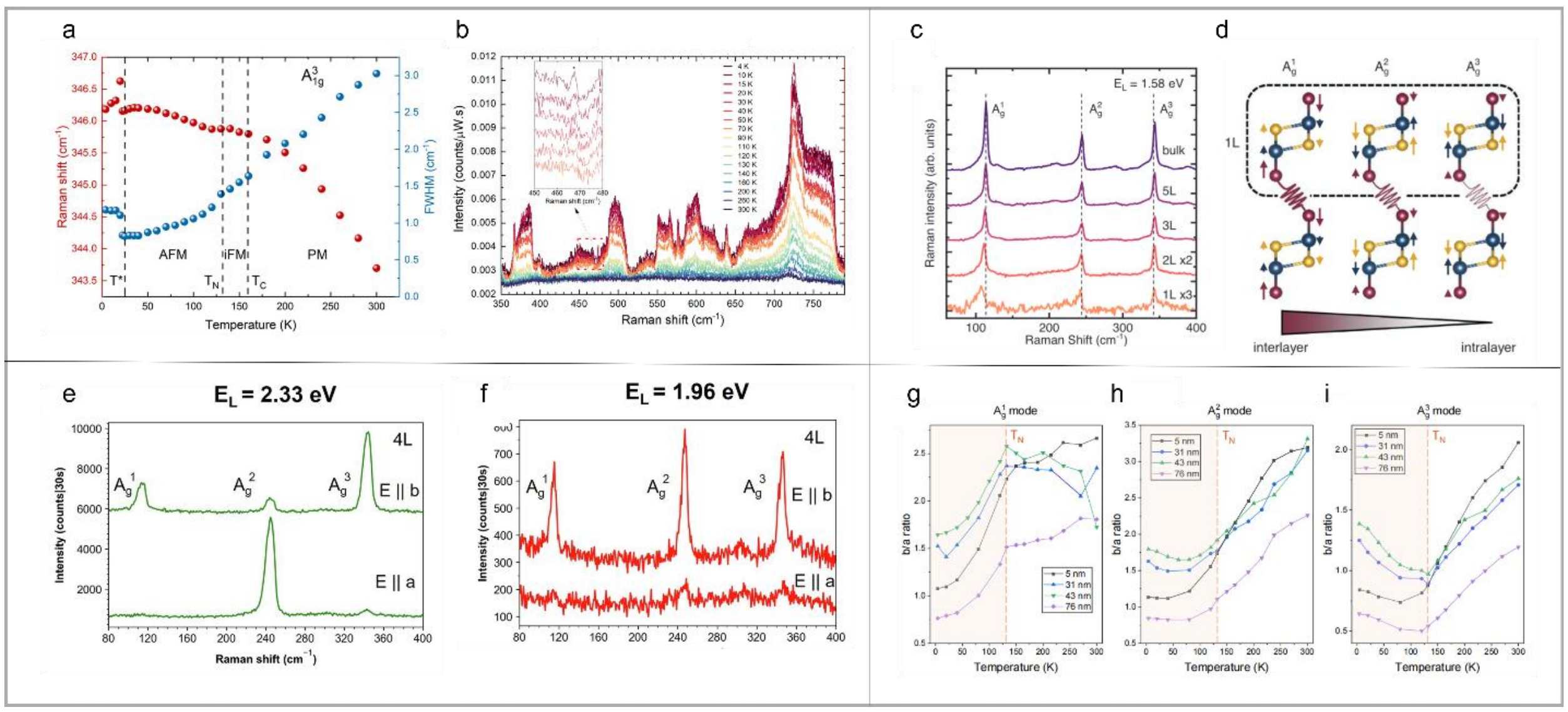

**Figure 3.** (a) The linewidth (blue dots) and phonon energy (red dots) are plotted against the temperature. (b) Raman scattering spectra at various temperatures, both below and above a certain point. The inset highlights the peak marked with a star, which emerges below. Reproduced with permission. [48] Copyright 2023, American Physical Society. (c) Raman spectra dependent on layer thickness are presented from a single layer to bulk for resonant excitation. (d) Illustrates atomic movements in all three Raman-active Ag modes for a single CrSBr layer, with two layers stacked to demonstrate interlayer coupling. The $A_g^1$ and $A_g^2$ modes involve Br atoms that enhance vibrational interlayer coupling, whereas the $A_g^3$ mode shows minimal Br involvement, resulting in a low interlayer coupling. Reproduced with permission. [55] Copyright 2023, Wiley-VCH. Beyond the thickness effects, polarization-resolved Raman measurements revealed the anisotropic characteristics of lattice vibrations in CrSBr. (e) Displays all Raman modes under an excitation energy of EL = 2.33 eV, and (f) shows them at EL = 1.96 eV for polarization along the b-axis and a-axis. Reproduced with permission. [58] Copyright 2025, Springer Nature. (g-i) Illustrates the temperature-dependent b/a ratio of Raman modes $A_g^1, A_g^2$ and $A_g^3$ under 1.96 eV excitation for varying layer numbers. Each Raman mode exhibited a unique behavior near $T_N$ across a broad thickness range. The orange-shaded region denotes the AFM phase, and the orange dashed line marks $T_N$=132 K. Reproduced with permission. [61] Copyright 2026, Springer Nature.

Temperature-dependent polarization measurements further underscore the role of excitonic effects in this system. Pronounced changes in the Raman anisotropy near the magnetic transition reveal a direct correlation between the magnetic ordering and optical selection rules **(Figure 3 g- i)** [61]. Combined analysis of the Raman and optical absorption data suggests that excitonic resonances modify electron–phonon interactions, thereby increasing the phonon sensitivity to the magnetic order. Consequently, the Raman response of CrSBr is governed not only by spin–phonon coupling but also by a complex interplay involving excitonic, electronic, and vibrational degrees of freedom. This integrated perspective situates CrSBr within a select group of layered magnetic semiconductors, where the magnetic order, lattice vibrations, and excitonic states are strongly interdependent. The transition from predominantly phononic features at elevated temperatures to coupled spin-phonon and exciton–phonon phenomena at low temperatures establishes a crucial framework for interpreting the optical properties discussed in the subsequent sections, where excitonic effects and magneto-optical responses are dominant.

## 5. Thickness-Dependent Excitonic Landscape and Magneto-Optical Response of CrSBr

The strong interplay among the electronic, magnetic, and excitonic degrees of freedom in CrSBr results in a complex and rich optical response that is distinct from that of conventional van der Waals semiconductors such as $CrI_3$, $CrCl_3$, and $LnSe_2$. In CrSBr, the excitonic properties are not solely determined by dielectric screening and quantum confinement but are significantly modulated by the magnetic order and interlayer electronic coupling. As the thickness increases from the monolayer limit to the bulk, the photoluminescence (PL) spectrum and its magnetic-field dependence undergo substantial changes, revealing a hierarchy of excitonic states that evolve with dimensionality [36, 62].

Initial PL studies on atomically thin CrSBr demonstrated robust excitonic emission at room temperature from 1L to trilayer (3L) samples, with emission peaks near 1.28–1.29 eV and absorption resonances of approximately 1.36–1.37 eV, corresponding to a Stokes shift of approximately 80 meV [17]. These transition energies align well with the theoretical electronic band structure predictions, confirming the semiconducting nature of CrSBr [63]. Temperature-dependent measurements showed the expected blue shift of excitonic emission upon cooling, which is characteristic of direct-gap semiconductors[64, 65]. However, deviations from this trend coincide with the onset of magnetic ordering, indicating a strong coupling between the excitonic states and the magnetic ground state [23].

Magneto-PL experiments revealed a pronounced thickness dependence of the magnetic field sensitivity. The monolayer CrSBr exhibited negligible magnetic-field effects on its PL spectrum, whereas the bilayer 2L CrSBr displayed significant field-induced spectral shifts and modifications in the excitonic lineshape **(Figure 4a–c)** [23]. This contrast highlights the critical role of interlayer electronic coupling in governing magneto-excitonic interactions and demonstrates that dimensionality fundamentally alters the coupling between excitons and the magnetic order.

At low temperatures (4 K), the PL spectra primarily consist of two bands, B1 and B2 **(Figure 4d)** [64]. The lower-energy B1 band, centered near 1.23 eV, dominates thicker flakes and bulk crystals but diminishes as the two-dimensional limit is approached. This band is attributed to impurity-related states, likely associated with sulfur and bromine vacancies [64, 66]. Conversely, the higher-energy B2 band persists across all thicknesses and is assigned to intrinsic exciton recombination [64, 66-68]. Whereas the 1 L and 2 L samples exhibited a single dominant excitonic feature (P1*), the thicker samples exhibited multiple excitonic transitions (P1, P2, P3, and P4) at lower energies. Notably, the P4 transition emerged from trilayer thickness onward and demonstrated exceptional magnetic field sensitivity, reflecting a strong coupling to interlayer magnetic correlations. The PL spectrum under an external magnetic field **B** for tri-layer and 15-layer CrSBr can be seen in **Figures 4e** and **4f** [64].

Magnetic correlation spectroscopy further reveals the complexity of the excitonic landscape [69]. Excitation power-dependent measurements confirmed the excitonic origin of most optical transitions through linear scaling. However, several emission lines exhibited saturation at higher powers, indicative of localized excitonic states confined within shallow trapping potentials[70]. Temperature-dependent studies have shown that these localized excitons vanish above approximately 16 K owing to thermal activation, permitting carriers to escape confinement and quenching the associated emission peaks [71, 72]. The magnetic-field-dependent data revealed significant spectral weight redistribution among excitonic transitions, emphasizing the interplay between magnetic correlations and exciton localization.

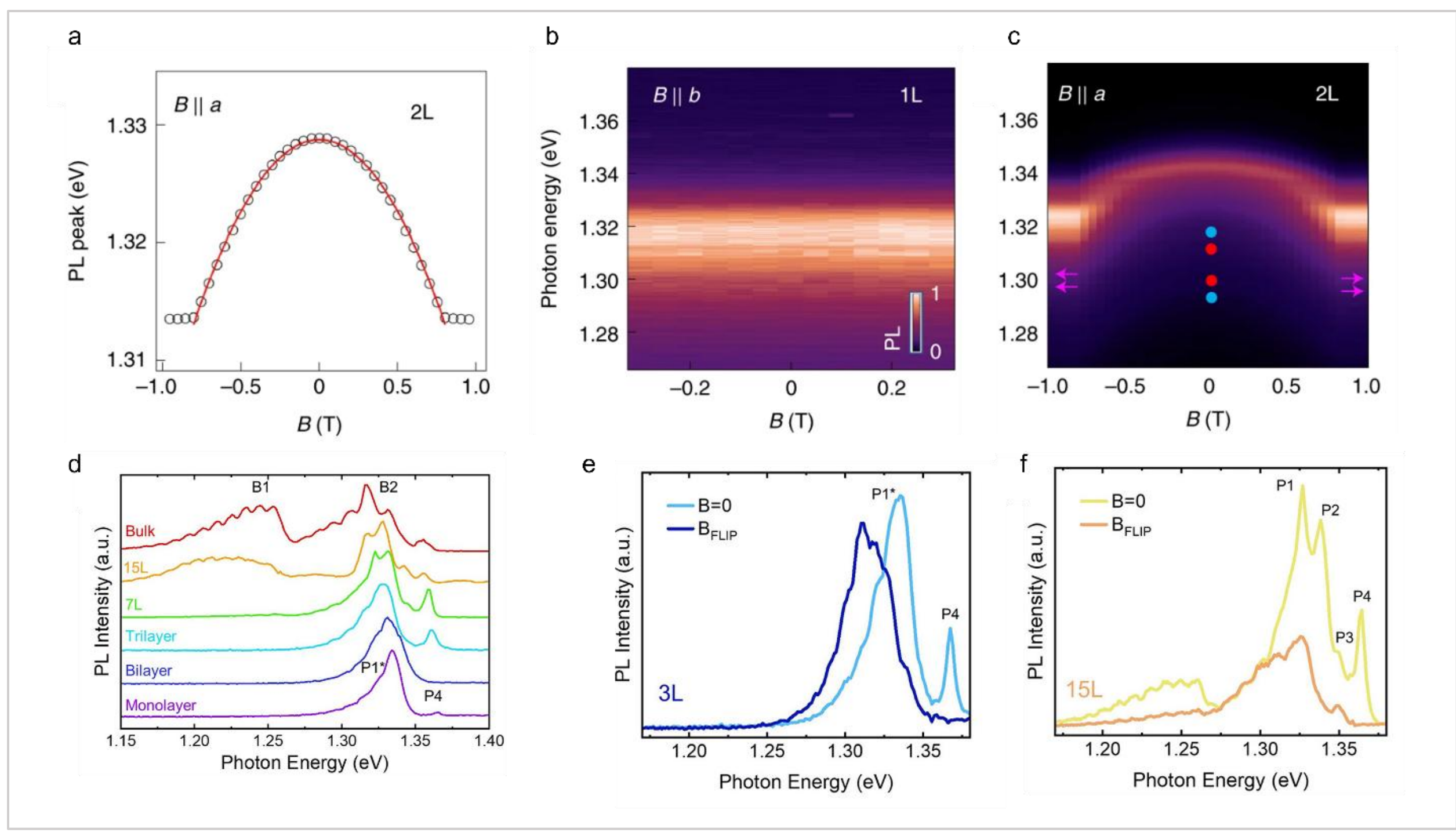

**Figure 4**. (a) Field-dependent photoluminescence peak position of bilayer CrSBr measured under an applied magnetic field oriented along the crystallographic a-axis. The red curve represents the fit to the experimental data. (b) PL spectra of monolayer CrSBr as a function of the external magnetic field applied along the a-axis. (c) PL spectra of BL CrSBr recorded with a magnetic field applied along the c-axis. Reproduced with permission. [23] Copyright 2021, Springer Nature. (d) Layer-dependent PL spectra of CrSBr flakes measured at 4 K. (e) and (f) PL peaks of tri-layer and 15-layer CrSBr, respectively, under an applied magnetic field at 4 K. Reproduced with permission. [64] Copyright 2023, American Chemical Society.

Additional insights have been gained from studies on hBN-encapsulated few-layer CrSBr. The monolayers displayed a broad excitonic feature near 1.34 eV, whereas the bilayers and thicker samples exhibited additional narrow emission lines with linewidths of a few meV. The behavior of the PL peaks in this study is shown in **Figure 5a** [73]. A high-energy excitonic state (Xh) appears in bilayers and splits into two distinct transitions (Xh and Xh') in trilayers and tetralayers, separated by approximately 5 μeV. These observations indicate a progressive reconstruction of the excitonic spectrum driven by an increase in interlayer coupling, reinforcing the strong correlation between dimensionality and optical response [74, 75].

Temperature-dependent photoluminescence measurements further confirmed that as the temperature increased, the excitons localized in shallow trapping potentials acquired sufficient thermal energy to escape confinement, leading to the disappearance of the corresponding emission peaks [76]. At elevated temperatures (~100 K), only selected emission peaks, such as XB, XD, and X3, persisted **(Figure 5b, c)**. The dependence of the emission peaks on the applied magnetic fields was also investigated, demonstrating that the excitonic emission of CrSBr is highly sensitive to magnetic fields. In the absence of a magnetic field, all emission peaks are visible **(Figure 5d,e),** whereas many peaks (P1, P2, P3, XD, X5, and X4) vanish upon applying magnetic fields of ±0.5 T [69].

A particularly remarkable recent development is the discovery of excitonic duality in CrSBr. Magneto-optical studies have revealed the coexistence of localized Frenkel-like excitons and delocalized Wannier-Mott excitons in the same material [77, 78]. This coexistence stems from the strongly anisotropic electronic structure of CrSBr, which offers a unique opportunity to simultaneously investigate two fundamentally distinct excitonic regimes. These exciton families exhibit markedly different responses to magnetic fields: Wannier-Mott excitons display pronounced spectral shifts and intensity redistribution, whereas Frenkel excitons remain largely localized with minimal field dependence. Complementary studies spanning monolayer to trilayer CrSBr revealed a striking odd-even layer dependence in the magneto-optical behavior; 1L and 3L samples exhibited similar characteristics, whereas 2L CrSBr displayed qualitatively distinct excitonic responses [3]. The representative magneto-PL and PL excitation (PLE) spectra of few-layer CrSBr acquired under tunable out-of-plane magnetic fields are shown in **Figure 5 f–h**. **Figure 5f** shows two-dimensional color maps of the PLE intensity as a function of excitation and detection energies for trilayer (3L) CrSBr, measured at magnetic fields ranging from 0 to 3.0 T. In the absence of an applied field, two distinct excitonic signatures were readily resolved: the high-energy B-exciton excitation band near 1.78 eV and a narrow A-series detection peak fixed at ~1.36 eV. With increasing magnetic field, the integrated PLE signal of the A′ subband (arising from bound or intervalley excitons bridging the B excitation manifold and A detection peak) increased drastically and monotonically, whereas the excitation and emission peak energies exhibited negligible magnetic-field-induced shifts.

Furthermore, the complementary stacked magneto-PL and magneto-PLE spectra shown in **Figure 5g** quantify the field-dependent evolution of the spectral intensity. The left-hand PL panel in **Figure 5g** shows the emission detected at the A exciton energy, revealing the progressive suppression of the A′ shoulder peak relative to the main A peak as the magnetic field increased.

The PLE in the right panel of **Figure 5g** shows the monitoring absorption at the A detection channel, showing pronounced amplification of the B-exciton excitation resonance with increasing external magnetic field. **Figure 5h** further shows the critical effect of layer thickness on the intrinsic PL peak shape. A primary source of peak profile reshaping between 1L, 2L and 3L studies.

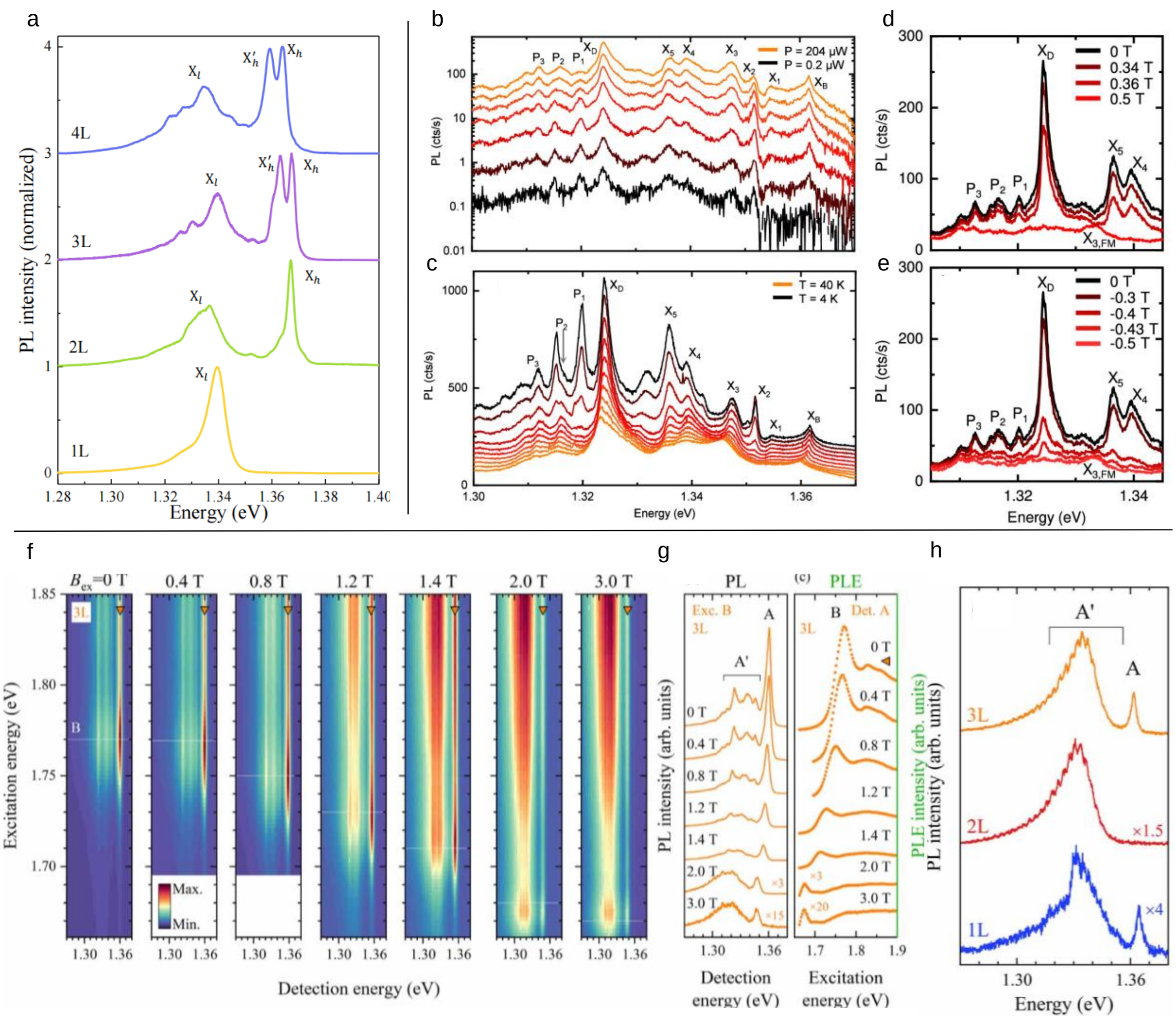

**Figure 5.** (a) Normalized PL spectra of CrSBr samples with thicknesses of 1-4 layers [73] Copyright 2025, arXiv. (b) Photoluminescence emission intensity as a function of excitation power ranging from 0.2 μW to 204 μW at 4.7 K with no applied magnetic field. (c) Temperature dependence of the PL emission measured at a constant excitation power of 10 μW in the absence of a magnetic field. (d) PL spectra of a CrSBr flake recorded under selected positive magnetic field strengths (Bb) applied along the crystallographic b-axis, showing a consistent decrease in the PL intensity with increasing field. (e) Corresponding PL spectra under selected negative magnetic fields along the b-axis, illustrating similar intensity reduction trends. Reproduced with permission. [69]

Copyright 2025, American Chemical Society. (f) False-color map of the PLE spectroscopy measured on 3L CrSBr around the A exciton for different values of magnetic field ranging from 0 to 3T. (g) PL and PLE spectra of 1L, 2L, and 3L CrSBr measured at 5 K under 2.41 eV excitation laser. (h) PL spectra of 3L CrSBr recorded at the B exciton resonance, demonstrating shifts in the B exciton energies with increasing applied magnetic field. Reproduced with permission. [3] Copyright 2025, IOP Science.

For 1L and 3L, the A' low-energy shoulder is weak and broad, with the leading A exciton peak, while 2L exhibits a moderately intensified A' feature, but no A peak is observed in the spectrum. Thickness-dependent quantum confinement controls interlayer orbital hybridization and exciton binding energies, modifying the relative oscillator strength of the A′ sub-exciton state without shifting that of the central exciton state. Collectively, this study demonstrates that the extrinsic measurement conditions and intrinsic material dimensionality dominate the variable PL peak line shapes and intensity ratios [3].

The evolution of narrow A and A' excitonic resonances under a magnetic field highlights that dimensionality not only shapes the excitonic spectrum but also fundamentally alters the exciton-magnetic-order interactions. Collectively, these studies demonstrate the progressive evolution of the excitonic landscape from relatively simple emission in the monolayer limit to a complex manifold of magnetically sensitive excitonic states in multilayer and bulk CrSBr. The strong dependence of excitonic transitions on thickness, magnetic ordering, interlayer coupling, and external magnetic fields underscores the intimate relationship between dimensionality and optical response, establishing excitons as highly sensitive probes of correlated electronic and magnetic phenomena in layered magnetic semiconductors.

## 6. Future Perspectives, Challenges, and Emerging Technological Applications

CrSBr is a remarkable early vdW magnet owing to its robust air stability and antiferromagnetic ordering up to ~132 K in bulk and ~140 K in ultrathin layers. Its structural anisotropy, layer-dependent magnetism, strong spin–phonon coupling, rich excitonic features, and magneto-optical response make it an ideal platform for studying spin, lattice, charge, and exciton interactions in reduced dimensions. Despite these significant advances, several critical challenges remain. A primary objective in the field of two-dimensional magnetism is to achieve magnetic ordering at or above room temperature. For CrSBr, promising strategies to achieve this goal include strain engineering, chemical substitution, electrostatic doping, and heterostructure design, which offer avenues for modulating exchange interactions and magnetic anisotropy. Tailoring super-exchange pathways through compositional modifications or leveraging proximity effects in hybrid heterostructures may enhance the magnetic ordering temperatures and stabilize novel magnetic phases.

Another important direction is the development of optically addressable magnetic functionality. Given the strong sensitivity of excitonic transitions to the magnetic order in CrSBr, optical spectroscopy has emerged as a powerful noninvasive probe of spin configurations in atomically thin layers. This capability opens pathways for integrating magnetic and optical functionalities within a single material platform, wherein magnetic states can be monitored and potentially manipulated using their excitonic signatures. Such integration holds promise for low-power information processing technologies and advances in the study of light–matter interactions in magnetic semiconductors. Beyond isolated flakes, the incorporation of CrSBr into engineered heterostructures and optical cavities represents a promising frontier. Recent discoveries of robust exciton–magnon and magneto-excitonic interactions suggest that CrSBr may host novel hybrid quasiparticles when coupled with confined photonic modes. The realization of magnetically tunable exciton-polaritons and related collective excitations could enable the exploration of nonequilibrium quantum phenomena and strongly coupled light–matter systems. Concurrently, continued progress in material synthesis, interface engineering, and device fabrication is essential to translate the intrinsic physical properties of CrSBr into viable technological applications.

In summary, CrSBr occupies a distinctive position at the nexus of two-dimensional magnetism, excitonic physics, and magneto-optics. Ongoing efforts to deepen the understanding and control of its coupled electronic, magnetic, vibrational, and optical properties are anticipated to uncover new physical phenomena and broaden the scope of applications enabled by this versatile, layered magnetic semiconductor.

## Acknowledgments

This work is supported by the European Union's Marie Skłodowska-Curie Actions under the Horizon Europe research and innovation programme (Grant No. 101204286). M.R.Molas gratefully acknowledges support from the National Science Centre, Poland (Grant No. 2022/46/E/ST3/00166 and 2023/50/O/ST3/00310).

**Author Contributions**

Muhammad Aftab and Chinmay Kumar Mohanty wrote the original draft of the manuscript. Zain Ashfaq and Warisha Mehmood contributed to manuscript editing and visualization. Xiaoli Wang and Clément Faugeras reviewed and edited the manuscript. Wajid Ali and Maciej R. Molas supervised the work and acquired funding. All authors reviewed and approved the final version of the manuscript.

**Data Availability**

Data will be made available on request.

**References**

1. C. Gong, L. Li, Z. Li et al., "Discovery of intrinsic ferromagnetism in two-dimensional van der Waals crystals," *Nature.* 546 (2017): 265-269
2. B. Huang, G. Clark, E. Navarro-Moratalla et al., "Layer-dependent ferromagnetism in a van der Waals crystal down to the monolayer limit," *Nature.* 546 (2017): 270-273
3. I. Antoniazzi, Ł. Kipczak, B. Camargo et al., "Magneto-Excitonic Duality From Monolayer to Trilayer CrSBr," *2D Materials.* 13 (2025): 015032
4. R. Cheng, L. Yin, Y. Wen et al., "Ultrathin ferrite nanosheets for room-temperature two-dimensional magnetic semiconductors," *Nature Communications.* 13 (2022): 5241
5. E. Wang, X. Zou, "Moiré bands in twisted trilayer black phosphorene: effects of pressure and electric field," *Nanoscale.* 14 (2022): 3758-3767

6. Q. H. Wang, A. Bedoya-Pinto, M. Blei et al., "The magnetic genome of two-dimensional van der Waals materials," *ACS Nano.* 16 (2022): 6960-7079
7. E. J. Telford, A. H. Dismukes, K. Lee et al., "Layered antiferromagnetism induces large negative magnetoresistance in the van der Waals semiconductor CrSBr," *Advanced Materials.* 32 (2020): 2003240
8. Y. Liu, W. Wang, H. Lu et al., "The environmental stability characterization of exfoliated few-layer $CrXTe_3$ (X= Si, Ge) nanosheets," *Applied Surface Science.* 511 (2020): 145452
9. M. Galbiati, V. Zatko, F. Godel et al., "Very long-term stabilization of a 2d magnet down to the monolayer for device integration," *ACS Applied Electronic Materials.* 2 (2020): 3508-3514
10. J. T. Gish, D. Lebedev, T. K. Stanev et al., "Ambient-stable two-dimensional $CrI_3$ via organic-inorganic encapsulation," *ACS Nano.* 15 (2021): 10659-10667
11. S. Jiang, L. Li, Z. Wang et al., "Controlling magnetism in 2D $CrI_3$ by electrostatic doping," *Nature Nanotechnology.* 13 (2018): 549-553
12. W. Jin, H. H. Kim, Z. Ye et al., "Raman fingerprint of two terahertz spin wave branches in a two-dimensional honeycomb Ising ferromagnet," *Nature Communications.* 9 (2018): 5122
13. S. Li, Z. Ye, X. Luo et al., "Magnetic-Field-Induced Quantum Phase Transitions in a van der Waals Magnet," *Physical Review X.* 10 (2020): 011075
14. Ł. Kipczak, A. Karmakar, M. Grzeszczyk et al., "Resonant Raman scattering of few layers CrBr3," *Scientific Reports.* 14 (2024): 7484
15. Ł. Kipczak, Z. Chen, M. Grzeszczyk et al., "Interplay between charge transfer and magnetic proximity effects in $WSe_2/CrCl_3$ heterostructures," *Nanoscale Horizons.* 10 (2025): 2465-2474
16. F. Dirnberger, J. Quan, R. Bushati et al., "Magneto-optics in a van der Waals magnet tuned by self-hybridized polaritons," *Nature.* 620 (2023): 533-537
17. T. Wang, D. Zhang, S. Yang et al., "Magnetically-dressed CrSBr exciton-polaritons in ultrastrong coupling regime," *Nature Communications.* 14 (2023): 5966
18. C. Ye, C. Wang, Q. Wu et al., "Layer-dependent interlayer antiferromagnetic spin reorientation in air-stable semiconductor CrSBr," *ACS Nano.* 16 (2022): 11876-11883

19. C. Boix-Constant, S. Mañas-Valero, A. M. Ruiz et al., "Probing the spin dimensionality in Single-layer CrSBr van der Waals heterostructures by magneto-transport measurements," *Advanced Materials.* 34 (2022): 2204940
20. X. Bo, F. Li, X. Xu et al., "Calculated magnetic exchange interactions in the van der Waals layered magnet CrSBr," *New Journal of Physics.* 25 (2023): 013026
21. O. Göser, W. Paul, H. Kahle, "Magnetic properties of CrSBr," *Journal of Magnetism and Magnetic Materials.* 92 (1990): 129-136
22. K. Lee, A. H. Dismukes, E. J. Telford et al., "Magnetic order and symmetry in the 2D semiconductor CrSBr," *Nano Letters.* 21 (2021): 3511-3517
23. N. P. Wilson, K. Lee, J. Cenker et al., "Interlayer electronic coupling on demand in a 2D magnetic semiconductor," *Nature Materials.* 20 (2021): 1657-1662
24. F. Tabataba-Vakili, H. P. Nguyen, A. Rupp et al., "Doping-control of excitons and magnetism in few-layer CrSBr," *Nature Communications.* 15 (2024): 4735
25. M. Liebich, M. Florian, N. Nilforoushan et al., "Controlling Coulomb correlations and fine structure of quasi-one-dimensional excitons by magnetic order," *Nature Materials* 24 (2025): 384-390
26. H. Li, Y. Yang, Z. Xia et al., "Stacking effects on magnetic, vibrational, and optical properties of CrSBr bilayers," *Physical Review B.* 111 (2025): 125411
27. V. Porée, A. Zobelli, A. Pawbake et al., "Resonant x-ray spectroscopies on CrSBr: Probing the electronic structure through chromium excitations," *Physical Review B.* 112 (2025): 125103
28. C. W. Cho, A. Pawbake, N. Aubergier et al., "Microscopic parameters of the van der Waals CrSBr antiferromagnet from microwave absorption experiments," *Physical Review B.* 107 (2023): 094403
29. S. A. López-Paz, Z. Guguchia, V. Y. Pomjakushin et al., "Dynamic magnetic crossover at the origin of the hidden-order in van der Waals antiferromagnet CrSBr," *Nature Communications.* 13 (2022): 4745
30. J. Beck, "Über chalkogenidhalogenide des chroms synthese, kristallstruktur und magnetismus von chromsulfidbromid, CrSBr," *Zeitschrift für anorganische und allgemeine Chemie.* 585 (1990): 157-167

31. N. Liu, C. Wang, Y. Zhang et al., "Intralayer strain-tuned interlayer magnetism in bilayer CrSBr," *Physical Review B.* 109 (2024): 214422
32. M. E. Ziebel, M. L. Feuer, J. Cox et al., "CrSBr: An Air-Stable, Two-Dimensional Magnetic Semiconductor," *Nano Letters.* 24 (2024): 4319-4329
33. Y. Guo, B. Wang, X. Zhang et al., "Magnetic two-dimensional layered crystals meet with ferromagnetic semiconductors," *InfoMat.* 2 (2020): 639-655
34. S. Badola, A. Pawbake, B. Wu et al., "van der Waals CrSBr Alloys with Tunable Magnetic and Optical Properties," *Nano Letters.* 26 (2026): 773-779
35. Z. Wang, M. Gibertini, D. Dumcenco et al., "Determining the phase diagram of atomically thin layered antiferromagnet $CrCl_3$," *Nature Nanotechnology.* 14 (2019): 1116-1122
36. H. Wang, J. Qi, X. Qian, "Electrically tunable high Curie temperature two-dimensional ferromagnetism in van der Waals layered crystals," *Applied Physics Letters.* 117 (2020):
37. T. J. Williams, A. A. Aczel, M. D. Lumsden et al., "Magnetic correlations in the quasi-two-dimensional semiconducting ferromagnet $CrSiTe_3$," *Physical Review B.* 92 (2015): 144404
38. G. Kresse, J. Furthmüller, "Efficiency of ab-initio total energy calculations for metals and semiconductors using a plane-wave basis set," *Computational Materials Science.* 6 (1996): 15-50
39. A. Pawbake, T. Pelini, I. Mohelsky et al., "Magneto-Optical Sensing of the Pressure-Driven Magnetic Ground States in Bulk CrSBr," *Nano Letters.* 23 (2023): 9587-9593
40. E. Henríquez-Guerra, A. M. Ruiz, M. Galbiati et al., "Strain Engineering of Magnetoresistance and Magnetic Anisotropy in CrSBr," *Advanced Materials.* n/a 2506695
41. M. A. Tschudin, D. A. Broadway, P. Siegwolf et al., "Imaging nanomagnetism and magnetic phase transitions in atomically thin CrSBr," *Nature Communications.* 15 (2024): 6005
42. C. Chen, Y. Liu, H. Lu et al., "Chemical vapor deposition growth of continuous monolayer antiferromagnetic CrOCl films," *Nature Communications.* 16 (2025): 11178
43. B.-Y. Yu, Y. Sun, X. Cao et al., "Strain effects on the lattice thermal conductivity of monolayer CrOCl: A first-principles study," *Materials Today Communications.* 38 (2024): 107665
44. Y. Yang, Y. Zhou, G. Zhang et al., "Thickness-Dependent Heat Dissipation in CrOCl Heat-Escaping Channel," *Advanced Functional Materials.* 35 (2025): 2412469

45. Y. Xu, A. Ray, Y. T. Shao et al., "Coexisting ferromagnetic–antiferromagnetic state in twisted bilayer $CrI_3$," *Nature Nanotechnology.* 17 (2022): 143-147
46. A. Ebrahimian, A. Dyrdał, A. Qaiumzadeh, "Control of magnetic states and spin interactions in bilayer $CrCl_3$ with strain and electric fields: an ab initio study," *Scientific Reports.* 13 (2023): 5336
47. F. Yao, V. Multian, Z. Wang et al., "Multiple antiferromagnetic phases and magnetic anisotropy in exfoliated $CrBr_3$ multilayers," *Nature Communications.* 14 (2023): 4969
48. A. Pawbake, T. Pelini, N. P. Wilson et al., "Raman scattering signatures of strong spin-phonon coupling in the bulk magnetic van der Waals material CrSBr," *Physical Review B.* 107 (2023): 075421
49. S. Sahu, A. Hashemi, M. Ghorbani-Asl et al., "Robust phonon engineering and symmetry-selective lattice dynamics in CrSBr1–xClx," *Journal of Physics: Materials.* 9 (2026): 035004
50. R. S. Das, Y. K. Agrawal, "Raman spectroscopy: Recent advancements, techniques and applications," *Vibrational Spectroscopy.* 57 (2011): 163-176
51. R. R. Jones, D. C. Hooper, L. Zhang et al., "Raman Techniques: Fundamentals and Frontiers," *Nanoscale Research Letters.* 14 (2019): 231
52. W. Min, J.-X. Cheng, Y. Ozeki, "Theory, innovations and applications of stimulated Raman scattering microscopy," *Nature Photonics.* 19 (2025): 803-816
53. Ł. Kipczak, T. Woźniak, C. Mohanty et al., *Strong Spin-Lattice Interaction in Layered Antiferromagnetic $CrCl_3$*. 2026.
54. Z. Jiang, P. Wang, J. Xing et al., "Screening and Design of Novel 2D Ferromagnetic Materials with High Curie Temperature above Room Temperature," *ACS Applied Materials & Interfaces.* 10 (2018): 39032-39039
55. K. Torres, A. Kuc, L. Maschio et al., "Probing Defects and Spin-Phonon Coupling in CrSBr via Resonant Raman Scattering," *Advanced Functional Materials.* 33 (2023): 2211366
56. C. Lee, H. Yan, L. E. Brus et al., "Anomalous lattice vibrations of single- and few-layer $MoS_2$," *ACS Nano.* 4 (2010): 2695-2700
57. A. Molina-Sánchez, L. Wirtz, "Phonons in single-layer and few-layer $MoS_2$ and $WS_2$," *Physical Review B.* 84 (2011): 155413

58. P. Mondal, D. I. Markina, L. Hopf et al., "Raman polarization switching in CrSBr," *npj 2D Materials and Applications.* 9 (2025): 22

59. T. Halenkovič, P. Němec, V. Nazabal, "Cluster-based DFT modeling of Raman vibrations in tetrahedral $GeS_2$ and $GeSe_2$ amorphous chalcogenides," *Scientific Reports.* 16 (2026): 10009

60. Z. Chen, W. Hwang, M. Cho et al., "In-plane optical and electrical anisotropy in low-symmetry layered GeS microribbons," *NPG Asia Materials.* 14 (2022): 41

61. D. I. Markina, P. Mondal, L. Krelle et al., "Interplay of vibrational, electronic, and magnetic states in CrSBr," *npj Quantum Materials.* (2026):

62. C. Wang, X. Zhou, L. Zhou et al., "A family of high-temperature ferromagnetic monolayers with locked spin-dichroism-mobility anisotropy: MnNX and CrCX (X = Cl, Br, I; C = S, Se, Te)," *Science Bulletin.* 64 (2019): 293-300

63. S. Reimers, L. Odenbreit, L. Šmejkal et al., "Direct observation of altermagnetic band splitting in CrSb thin films," *Nature Communications.* 15 (2024): 2116

64. F. Marques-Moros, C. Boix-Constant, S. Mañas-Valero et al., "Interplay between Optical Emission and Magnetism in the van der Waals Magnetic Semiconductor CrSBr in the Two-Dimensional Limit," *ACS Nano.* 17 (2023): 13224-13231

65. K. Lin, X. Sun, F. Dirnberger et al., "Strong Exciton–Phonon Coupling as a Fingerprint of Magnetic Ordering in van der Waals Layered CrSBr," *ACS nano.* 18 (2024): 2898-2905

66. J. Klein, Z. Song, B. Pingault et al., "Sensing the Local Magnetic Environment through Optically Active Defects in a Layered Magnetic Semiconductor," *ACS Nano.* 17 (2023): 288-299

67. J. Klein, B. Pingault, M. Florian et al., "The Bulk van der Waals Layered Magnet CrSBr is a Quasi-1D Material," *ACS Nano.* 17 (2023): 5316-5328

68. J. Cenker, S. Sivakumar, K. Xie et al., "Reversible strain-induced magnetic phase transition in a van der Waals magnet," *Nature Nanotechnology.* 17 (2022): 256-261

69. L. Krelle, R. Tan, D. Markina et al., "Magnetic Correlation Spectroscopy in CrSBr," *ACS Nano.* 19 (2025): 33156-33163

70. T.-X. Qian, J. Zhou, T.-Y. Cai et al., "Anisotropic electron-hole excitation and large linear dichroism in the two-dimensional ferromagnet CrSBr with in-plane magnetization," *Physical Review Research.* 5 (2023): 033143

71. T. Godde, D. Schmidt, J. Schmutzler et al., "Exciton and trion dynamics in atomically thin ${\mathrm{MoSe}}_{2}$ and ${\mathrm{WSe}}_{2}$: Effect of localization," *Physical Review B.* 94 (2016): 165301

72. X. Huang, Z. Song, Y. Gao et al., "Intrinsic Localized Excitons in $MoSe_2$/CrSBr Heterostructure," *Advanced Materials.* 37 (2025): 2413438

73. X. Wu, J. Chen, M. Gu et al. *Magnetic order dependent photoluminescence from high energy excitons in hBN protected few-layer CrSBr*. **2025** [arXiv:2507.23301 DOI: 10.48550/arXiv.2507.23301.

74. E. J. Telford, A. H. Dismukes, R. L. Dudley et al., "Coupling between magnetic order and charge transport in a two-dimensional magnetic semiconductor," *Nature Materials.* 21 (2022): 754-760

75. Z. Sun, C. Hong, Y. Chen et al., "Resolving and routing magnetic polymorphs in a 2D layered antiferromagnet," *Nature Materials.* 24 (2025): 226-233

76. R. Komar, A. Łopion, M. Goryca et al., "Colossal magneto-excitonic effects in 2D van der Waals magnetic semiconductor CrSBr," *arXiv e-prints.* (2024): arXiv: 2409.00187

77. M. Śmiertka, M. Rygała, K. Posmyk et al., "Distinct magneto-optical response of Frenkel and Wannier excitons in CrSBr," *Nature Communications.* 17 (2026): 1777

78. I. M. Pratap Chandra Adak, Suvodeep Paul, Caleb Heuvel-Horwitz, Bikash Das, Vitali Kozlov, Kseniia Mosina, Arun Ramanathan, Xavier Roy, Zdeněk Sofer, Tian Zhong, Akashdeep Kamra, Arno Thielens, Andrea Alù, Vinod M. Menon, "Microwave-to-optical transduction using magnon-exciton coupling in a layered antiferromagnet," *arXiv.* (2026): 2604.03441